\documentclass[aps,pra,showpacs,twocolumn,superscriptaddress]{revtex4}

\usepackage{graphicx,color}
\usepackage{amsmath,amsthm,amsfonts,amssymb,bm}
\usepackage{times}
\usepackage{epsf}
\usepackage[colorlinks={true}]{hyperref}

\usepackage[T1]{fontenc}
\usepackage[utf8]{inputenc}

\hypersetup{citecolor={blue}, filecolor={blue}, linkcolor={blue}, urlcolor={blue}}
\begin{document}

\title{Sudden change of quantum discord for a system of two qubits}

\author{Jo\~{a}o P. G. Pinto}
\affiliation{Departamento de F\'{\i}sica, Universidade Federal de Ouro Preto, Ouro Preto, MG, 35400-000, Brazil}
\author{G\"{o}ktu\u{g} Karpat}
\affiliation{Faculdade de Ci\^encias, UNESP - Universidade Estadual Paulista, Bauru, SP, 17033-360, Brazil}
\author{Felipe F. Fanchini}
\affiliation{Faculdade de Ci\^encias, UNESP - Universidade Estadual Paulista, Bauru, SP, 17033-360, Brazil}

\begin{abstract}
It is known that quantum discord might experience a sudden transition in its dynamics when calculated for certain Bell-diagonal states (BDS) that are in interaction with their surroundings. We examine this phenomenon known as the sudden change of quantum discord, considering the case of two qubits independently interacting with dephasing reservoirs. We first numerically demonstrate that, for a class of initial states which can be chosen arbitrarily close to BDS, the transition is in fact not sudden, although it might numerically appear so if not studied carefully. Then, we provide an extension of this discussion covering the X-shaped density matrices. Our findings suggest that the transition of quantum discord might be sudden only for an highly idealized zero-measure subset of states within the set of all possible initial conditions of two qubits.
\end{abstract}


\maketitle

\section{Introduction}

In recent years, one of the most extensively studied topics of quantum information science has been the quantification and investigation of non-classical correlations that cannot be fully captured through entanglement measures \cite{discordreview}. Moreover, it has been shown that even completely separable mixed states might exhibit non-classical behaviour and offer computational speed-up in quantum information processes \cite{datta}. Although numerous quantities have been proposed to quantify such quantum correlations, that are more general than entanglement, quantum discord \cite{discord} has proven to be the most popular among these new measures. Quantum discord is present in almost all states in nature \cite{ferraro} and robust to sudden death in case of an interaction with a reservoir \cite{werlang}.

One of the most striking features of quantum discord is that it can experience a sudden transition when considering the dynamics of open quantum systems
\cite{suddenchange}. That is to say that the time derivative of quantum discord might become discontinuous at certain points during the time evolution of the system.
At such critical time instants, there exist sudden transitions between different dynamical behaviors and, even more curiously, it is possible for the time evolution of quantum discord to freeze and let it evolve independently of the destructive effects of the environment \cite{markovianfrozendiscord, nonmarkovianfrozendiscord}. This peculiar behaviour has intrigued the scientific community in the last few years since such a discontinuity in the time derivative of a quantum correlation measure had never been observed before.

In this paper, we explore the phenomenon of the sudden change of quantum discord for a system of two qubits that are independently interacting with dephasing reservoirs. Instead of choosing Bell-diagonal states (BDS), we consider slightly different initial conditions which are in the close vicinity of BDS. We first demonstrate that, contrary to the case of BDS, time derivative of quantum discord \textit{exists} at any instant for a class of states that can be chosen arbitrary close to BDS. Next, we extend our discussion to X-shaped density matrices. Our analysis suggests that quantum discord might exhibit a sudden change only for a zero-measure subset within the set of all two-qubit states, which leads us to conjecture that the phenomenon of sudden change almost never occurs in nature.

\section{Quantum Discord}

In this section, we introduce the definition of the quantum discord which has been originally proposed by Ollivier and Zurek as a measure of genuine quantum correlations \cite{discord}. The total amount of classical and quantum correlations present in a bipartite quantum system $\rho_{AB}$ is quantified with the help of the quantum mutual information,
\begin{equation}
I\left(\rho_{AB}\right)=S\left(\rho_{A}\right)+S\left(\rho_{B}\right)-S\left(\rho_{AB}\right),
\end{equation}
where $S\left(\rho\right)=-Tr\left(\rho\log_{2}\rho\right)$ is the von-Neumann entropy and the quantities $\rho_{A}=Tr_{B}\left(\rho_{AB}\right)$ and $\rho_{B}=Tr_{A}\left(\rho_{AB}\right)$ are the reduced density operators of the composite system $\rho_{AB}$, respectively. On the other hand, classical correlations in a bipartite quantum system $\rho_{AB}$ can be captured by \cite{classical},
\begin{equation}
J(\rho_{AB})=\max_{\{\Pi_{k}^{B}\}}\left\{ S\left(\rho_{A}\right)-\sum_{k}p_{k}S\left(\rho_{A|k}  \right)\right\}\label{cc}
\end{equation}
where $\{\Pi_{k}^{B}\}$ is a complete set of orthonormal projection operators that act only on the subsystem $B$, and $\rho_{A|k}=Tr_B(\Pi_{k}^{B}\rho_{AB}\Pi_{k}^{B})/p_k$ is the remaining state of the subsystem $A$ after obtaining the outcome $k$ with probability $p_k=Tr_{AB}(\Pi_{k}^{B}\rho_{AB}\Pi_{k}^{B})$ in the subsystem $B$. The local measurement operators $\Pi_{k}^{B}=|\pi_{k}\rangle\langle\pi_{k}| (k=1,2)$ can be easily constructed using the following quantum states:
\begin{align}
\vert\pi_{1}\rangle &= \cos\theta\vert0\rangle+e^{i\phi}\sin\theta\vert1\rangle, \\
\vert\pi_{2}\rangle &= \sin\theta\vert0\rangle-e^{i\phi}\cos\theta\vert1\rangle,
\end{align}
where $\theta \in [0,\pi]$ and $\phi \in [0, 2\pi]$. Then, quantum discord is defined as the difference between the quantum mutual information and the classical correlation,
\begin{equation}
D(\rho_{AB})=I(\rho_{AB})-J(\rho_{AB}).
\end{equation}

It is important to stress that it is difficult to obtain an closed analytical expression for quantum discord even in the case of two qubits due to the potentially complex optimization procedure in the definition of the classical information. We also note that quantum discord is not a symmetric quantity in general since its definition is dependent on the local subsystem the measurement is performed on.

\section{Non-Markovian Dephasing Model}

Having introduced the definition of quantum discord, we now turn our attention to the non-Markovian dephasing model that we intend to use in our investigation. Here, we consider a colored noise dephasing model first studied by Daffer et al. in Ref. \cite{model}. First, we assume that the dynamics is described by a master equation of the form
\begin{equation}
\dot{\rho}=K\mathcal{L}\rho, \label{master}
\end{equation}
where $K$ is a time-dependent integral operator that acts on the system as $K\phi=\int_0^t k(t-t')\phi(t')dt'$, $k(t-t')$ is a kernel function determining the type of memory in the environment, $\rho$ is the density operator of the principal system, and $\mathcal{L}$ is a Lindblad superoperator describing the open system dynamics as a result of the interaction between the principal system and the environment. This type of a master equation might arise if one considers a two-level quantum system that interacts with a reservoir having the properties of random telegraph signal noise. In order to analyze a master equation of the form of Eq. (\ref{master}), one might start with a time-dependent Hamiltonian,
\begin{equation}
H(t)=\hbar\sum_{k=1}^3\Gamma_k(t)\sigma_k,
\end{equation}
where $\sigma_k$ are the usual Pauli matrices and $\Gamma_k(t)$ are independent random variables obeying the statistics of a random telegraph signal. In particular, the random variables can be expressed as $\Gamma_k(t)=a_k n_k(t)$, where $n_k(t)$ has a Poisson distribution with a mean equal to $t/2\tau_k$ and $a_k$ is a coin-flip random variable possessing the values $\pm a_k$.

One can make use of the von Neumann equation, given by $\dot{\rho}=-(i/\hbar)[H,\rho]$, to obtain a solution for the density matrix of the two-level system having the form,
\begin{equation}
\rho(t)=\rho(0)-i \int_0^t\sum_k \Gamma_k(s)[\sigma_k,\rho(s)]ds. \label{isol}
\end{equation}
Then, substituting Eq. (\ref{isol}) back into the von Neumann equation and performing the stochastic averages, one ends up with
\begin{equation}
\dot{\rho}(t)=-\int_0^t\sum_k e^{-(t-t')/\tau_k}a_k^2 [\sigma_k,[\sigma_k,\rho(t')]]dt', \label{sol}
\end{equation}
where the memory kernel comes from the correlation functions of the random telegraph signals $\langle\Gamma_j(t)\Gamma_k(t')\rangle=a_k^2\exp(-|t-t'|/\tau_k)\delta_{jk}$.

At this point, we note that the model described above gives rise to a non-Markovian time evolution according to some recently introduced measures of non-Markovianity such as the ones proposed by Luo, Fu and Song (LFS) \cite{LFS} and Breuer, Laine and Pillo (BLP) \cite{BLP}.

Daffer et al. have also analyzed the conditions under which the dynamical evolution generated by Eq. (\ref{sol}) is completely positive and found that completely positivity is guaranteed when two of the $a_k$ are zero. Such a case corresponds to the physical situation where the noise only acts in one direction. Particularly, provided that the condition $a_3=a$ and $a_1=a_2=0$ holds, the dynamics experienced by the system is that of a dephasing channel with colored noise. In this case, the Kraus operators describing the reduced dynamics of the two-level system are give given by
\begin{align}
M_1 &= \sqrt{[1+\Lambda(\nu)]/2}I_2, \\
M_2 &= \sqrt{[1-\Lambda(\nu)]/2}\sigma_3,
\end{align}
where $I_2$ is the $2\times2$ identity matrix and the Kraus operators satisfy $\sum_k^2M_k^\dagger M_k=I$. Here, $\Lambda(\nu)=e^{-\nu}[\cos(\mu\nu)+\sin(\mu\nu)/\mu], \mu=\sqrt{(4a\tau)^2-1}$ and $\nu=t/2\tau$ is the dimensionless time. Since we intend to investigate the dynamics of two qubits which are independently interacting with identical colored dephasing environments, we can obtain the completely positive trace preserving map responsible for the time evolution of the two-qubit system as follows,
\begin{equation}
\rho_{AB}(t)=\sum_{i,j}(M_i^A\otimes M_j^B) \rho_{AB}(0) (M_i^A\otimes M_j^B)^\dagger,
\end{equation}
where $\rho_{AB}(0)$ is the initial state of the two-qubit system and the operators $M_i^A$ and $M_i^B$ act on the first and the second qubits, respectively.

\section{Main Results}

In this section, we first present our results related to the dynamics of quantum discord for a two-qubit system interacting with independent colored dephasing reservoirs. Then, we analyze possible sudden transitions between different measurement bases, which are used in the calculation of quantum discord, taking place during the time evolution of the system.

\begin{figure}[htbp]
\includegraphics[width=0.48\textwidth]{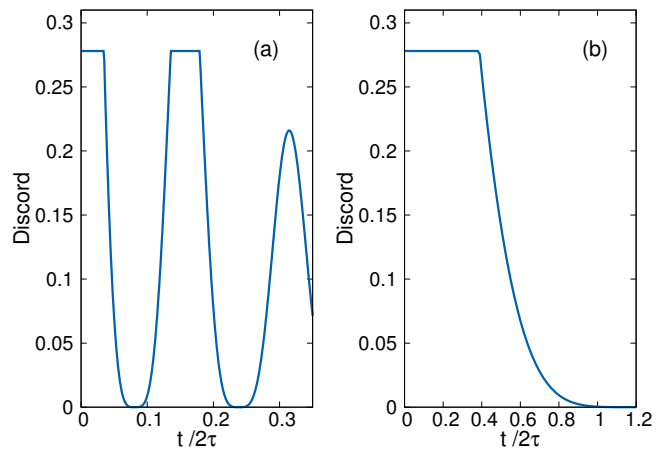}
\caption{(Color online) Dynamics of quantum discord for the BDS (given in Eq. (\ref{bds})) described by the parameters $c_1=1, c_2=-0.6$ and $c_3=0.6$ as a function of the dimensionless time $t/2\tau$ with (a) $a=1 s$ and $\tau=5 s$ and with (b) $a=1 s$, $\tau=0.5 s$. Note that while the dynamics presented in figure (a) has non-Markovian features, the process in figure (b) displays the Markovian limit of the considered dynamical quantum map.}
\end{figure}

\begin{figure*}[htbp]
\includegraphics[width=0.95\textwidth]{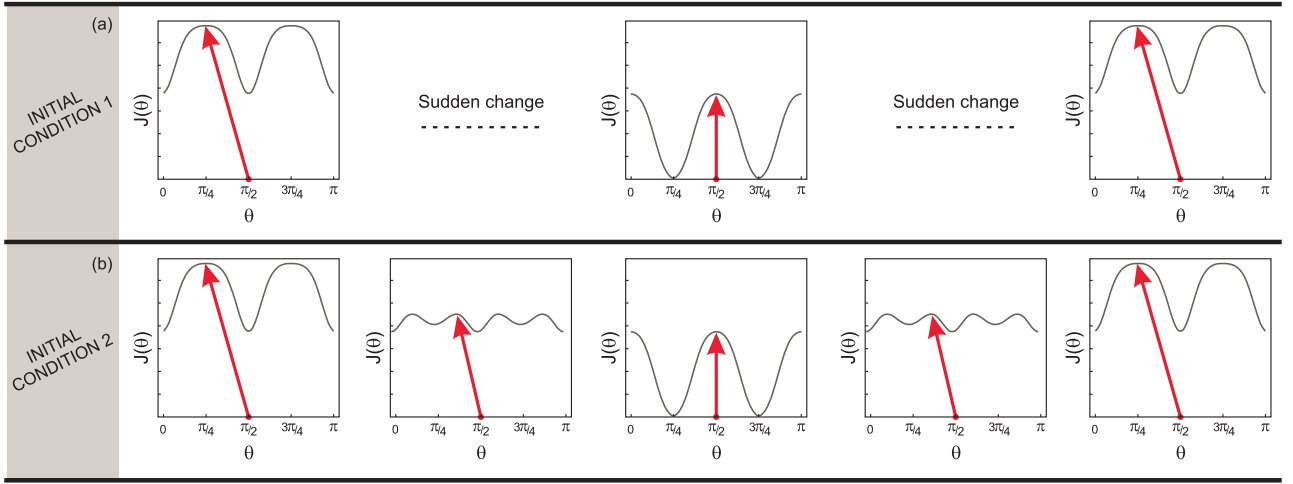}
\caption{(Color online) Possible values of classical correlation as a function of the angle $\theta$. (a) Initial Condition 1: Considering the BDS given by the parameters $c_1=1, c_2=-0.6, c_3=0.6$ as an initial condition, we see that the basis which maximizes quantum discord suddenly changes twice during the dynamics of the system, namely, from $\sigma_x$ to $\sigma_z$ and then back to $\sigma_x$. (b) Initial condition 2: Considering the initial state given by Eq. (\ref{ours}) with $c_1=1, c_2=-0.6, c_3=0.6$ and $\epsilon=0.02$, which is in the close vicinity of BDS, we observe that the transition from one basis to another is in fact not sudden and the time derivative of the quantum discord exists at any time instant throughout the dynamics.}
\end{figure*}

We will begin our investigation considering the initial states
\begin{equation}
\rho_{AB}=\frac{1}{4} \left(I_4 + \sum_{j=1}^3 c_j \sigma_j^A \otimes \sigma_j^B \right),\label{bds}
\end{equation}
where $I_4$ is the $4\times4$ identity matrix, $c_j$ are real numbers such that $0\leq |c_j| \leq1$, and the eigenvalues $(\lambda_k\geq0)$ of $\rho_{AB}$ are
\begin{align}
\lambda_{1,2} &= \frac{1}{4}(1\pm c_1\pm c_2-c_3),  \\
\lambda_{3,4} &= \frac{1}{4}(1\pm c_1\mp c_2+c_3).
\end{align}
The above states, known as BDS, form a tetrahedron with the four Bell states sitting in the extreme points, and have been widely used in the literature to demonstrate the sudden change of quantum discord in open quantum systems \cite{suddenchange,markovianfrozendiscord,nonmarkovianfrozendiscord,frozencond}.

At this point, it is important to mention that the family of states given in Eq. (\ref{bds}) is highly idealized. Considering the fact that a two-qubit density matrix is parameterized by a set of 15 real variables, it is impossible to prepare the desired state without any errors in a real world situation, independently of the experimental apparatus. For this reason, we will also consider the initial states that are very slightly different from BDS, having the form
\begin{equation}
\rho_{AB}=\frac{1}{4} \left(I_4 + \sum_{j=1}^3 c'_j \sigma_j^A \otimes \sigma_j^B + \epsilon(I_2^A \otimes \sigma_3^B + \sigma_3^A \otimes I_2^B) \right),\label{ours}
\end{equation}
where $c'_1=c_1-\epsilon, c'_2=c_2+\epsilon,c'_3=c_3$ and the parameter $\epsilon$ is considered to be a small number. Although the state given in Eq. (\ref{ours}) might also be considered ideal, as we will demonstrate shortly, it is sufficient for us to make our point about the sudden change of quantum discord.

We start our analysis presenting the dynamics of the BDS, described by the parameters $c_1=1, c_2=-0.6, c_3=0.6$, that is independently interacting with identical colored dephasing environments having $a=1 s$, $\tau=5 s$ in Fig. (1-a) and $a=1 s$, $\tau=0.5 s$ in Fig (1-b). It can be straightforwardly seen that whereas the quantum process studied in Fig. (1-a) exhibits revivals in quantum discord due to its non-Markovian behavior, Fig. (1-b) displays the Markovian limit of the same process.

Moreover, we notice that quantum discord experiences three sudden transitions for the non-Markovian dynamics \cite{nonmarkovianfrozendiscord}, as opposed to the case of Markovian dynamics, where only a single transition can be observed. Note that the transitions observed in Fig. (1-a) and Fig. (1-b) are truly sudden. For instance, the dynamics in the Markovian case is such that while $c_1$ and $c_2$ decay exponentially, $c_3$ remains constant in time. Since initially $|c_1|>|c_3|$, the sudden change occurs when $|c_1(t)|=|c_3|$ \cite{suddenchange}. At this critical time instant, the basis maximizing the classical correlation given in Eq. (\ref{cc}) changes abruptly, which results in a discontinuity in the time derivative of quantum discord. This intriguing phenomenon is known as the \textit{sudden change of the quantum discord} in the literature.

In the following, we will investigate the possibility of occurrence of such sudden transitions for the initial states given in Eq. (\ref{ours}), that is, for the states which can be chosen arbitrarily close to BDS. Here, we consider the non-Markovian dephasing dynamics, where the sudden transition occurs three times, to study the problem. To start with, Fig. (2-a) displays the behavior of the classical correlation as a function of the angle $\theta$ for the BDS described by the parameters $c_1=1$, $c_2=-0.6$, $c_3=0.6$.  In this case, it is possible to show analytically that the transition is actually sudden since the optimal measurement basis abruptly changes twice, first from $\sigma_x$ to $\sigma_z$ then back to $\sigma_x$. These two transition points correspond to the first and the second sudden change instants during the dynamics of the system as shown in Fig. (1-a). Next, in Fig. (2-b), we illustrate that the transition can be seen to be in fact continuous for the initial state given in Eq. (\ref{ours}) with the parameters $c_1=1$, $c_2=-0.6$, $c_3=0.6$, and $\epsilon=0.02$, provided that the dynamics is carefully investigated for short enough time intervals. In this case, we display the change of the optimal basis for some time instants during the transition. Contrary to the case of Fig. (2-a), the optimal angle giving the maximum, pointed by the red arrow, changes \textit{continuously} from $\pi/4$ to $\pi/2$ and then back to $\pi/4$. This shows that even a very slight deviation from the form BDS is enough to remove the discontinuity in the time derivative of the quantum discord. We also note that even when one chooses an $\epsilon$ that is not a small perturbation, one might still mistake a continuous change for a
sudden one, based on numerical calculations, since the transition between the bases takes place very quickly. Although we demonstrated this behavior for the non-Markovian dynamics, the same result can be obtained for the Markovian case.

Let us now consider the class of X-shaped density matrices with five real
coefficients in the computational basis,
\begin{eqnarray}
\rho=\left(\begin{array}{cccc}
\rho_{00} & 0 & 0 & \rho_{03}\\
0 & \rho_{11} & \rho_{12} & 0\\
0 & \rho_{12} & \rho_{22} & 0\\
\rho_{03} & 0 & 0 & \rho_{33}
\end{array}\right).
\end{eqnarray}
For such density matrices, Chen et al. have obtained the conditions on the matrix elements for optimal measurement basis to be $\sigma_x$ or $\sigma_z$ \cite{conditions}. In particular, they have shown that, provided we assume $|\rho_{12}+\rho_{03}| \geq |\rho_{12}-\rho_{03}|$, the optimal measurement for quantum discord is given by
$\sigma_z$ if $(|\rho_{12}|+|\rho_{03}|)^2\leq(\rho_{00}-\rho_{11})(\rho_{33}-\rho_{22})$, and by $\sigma_x$ if $|\sqrt{\rho_{00}\rho_{33}}-\sqrt{\rho_{11}\rho_{22}}|\leq|\rho_{12}|+|\rho_{03}|$.
Note that there exist some states that do not satisfy neither of the conditions. The optimal basis for these states will lie on the $x$-$z$ plane being a combination of $\sigma_x$ and $\sigma_z$ \cite{conditions}. Looking at the conditions, it is straightforward to observe that the two expressions coincide, with inequalities having opposite directions, when we have $\rho_{00}\rho_{22}=\rho_{11}\rho_{33}$. This constraint guarantees the absence of a small region of parameters where none of the inequalities are satisfied. Assuming a possible change of basis between $\sigma_x$ and $\sigma_z$, $\rho$ requires to satisfy this constraint for the transition to be truly sudden. For instance, suppose that a pure dephasing channel acts on the system $\rho$. Since diagonal elements are untouched by the process, one can immediately tell in this case whether a possible transition between $\sigma_x$ and $\sigma_z$ will be truly sudden or not. Actually, this also helps us understand why we do not have sudden change in our numerical example presented earlier unless $\epsilon=0$. Consequently, as a result of the restriction given by $\rho_{00}\rho_{22}=\rho_{11}\rho_{33}$, the dimension of the parameter space for the states exhibiting sudden change turns out to be smaller than the dimension of general real X-shaped states. This observation strengths our doubt that the sudden change might occur only for a zero-measure subset of states.

\section{Conclusion}

The phenomenon of sudden change of quantum discord has intrigued the scientific community since it was first observed for BDS interacting with independent dephasing environments \cite{suddenchange}. In this report, we have first numerically shown that, where the considered states are lying in the close vicinity of BDS, the transitions between different dynamical trends of quantum discord are actually continuous. In other words, the time derivative of quantum discord exists at any time instant. We have then extended our discussion to the X-shaped states having real parameters. Moreover, we have also tested the possibility of occurrence of sudden changes for $10^6$ randomly chosen X-shaped states and found no sign of such a dynamical behavior. We also emphasize that one should be careful about any claims of sudden transitions by looking at plots of quantum discord based on numerical calculations. In some cases, although it might seem like one has sudden change, the transitions can turn out to be continuous when analyzed for small enough time intervals. Considering that the sudden change phenomenon has only been reported
for X-shaped states to this date, our results lead us to conjecture that the change of quantum discord might in fact be sudden only for a zero-measure subset of states within the set of all possible initial conditions of two qubits.

\begin{acknowledgments}
JPGP is supported by Coordination for the Improvement of Higher Education Personnel (CAPES), GK is supported by National Counsel of Technological and Scientific Development (CNPQ), and FFF is supported by S\~{a}o Paulo Research Foundation (FAPESP) under grant number 2012/50464-0, and by the National Institute for Science and Technology of Quantum Information (INCT-IQ) under process number 2008/57856-6.
\end{acknowledgments}

\end{document}